\documentclass[twocolumn]{jpsj3}
\usepackage[usenames]{color}
\def\v#1{\mib #1}

\newcommand{\aver}[1]{\left\langle {#1} \right\rangle}
\def\Nc{N_{\rm c}}
\def\EHal{{\tilde{E}}}
\def\Jeff{J_{\rm eff}}
\def\Keff{K_{\rm eff}}
\def\H{{\cal H}}

\def\deltaa{{\delta_{\rm A}}}
\def\deltaaa{{\delta_{\rm A}^2}}
\def\deltab{{\delta_{\rm B}}}

\def\theenumi{{\roman{enumi}}} 

\def\runauthor{K. {\sc Hida}, K. {\sc Takano}, and  H. {\sc Suzuki}}
\title
{
Haldane Phases and Ferrimagnetic Phases with Spontaneous Translational Symmetry Breakdown in Distorted Mixed Diamond Chains with Spins $1$ and $1/2$}
\author
{
Kazuo {\sc Hida}\thanks{E-mail address: hida@phy.saitama-u.ac.jp}, 
Ken'ichi {\sc Takano}$^{1}$, and  
Hidenori {\sc Suzuki}$^{1}$\thanks{Present address: Department of Physics, College of Humanities and Sciences, Nihon University, Setagaya-ku, Tokyo 156-8550}}

\inst
{Division of Material Science, Graduate School of Science and Engineering, \\ Saitama University, Saitama, Saitama, 338-8570\\ 
$^{1}$Toyota Technological Institute, 
Tenpaku-ku, Nagoya 468-851}

\recdate
{
\today
}

\abst
{
The ground states of two types of distorted mixed diamond chains with spins $1$ and $1/2$ are investigated  using exact diagonalization, DMRG, and mapping onto low-energy effective models. 
In the undistorted case, the ground state consists of an array of independent spin-1 clusters separated by singlet dimers. 
The lattice distortion induces an  effective interaction between cluster spins. 
When this effective interaction is antiferromagnetic, several Haldane phases appear with or without spontaneous translational symmetry  breakdown (STSB). 
The transition between the Haldane phase without STSB and that with $(n+1)$-fold STSB ($n$ = 1, 2, and 3) belongs to the same universality class as the $(n+1)$-clock model. 
In contrast, when the effective interaction is ferromagnetic, the quantized and partial ferrimagnetic phases appear with or without STSB. 
An effective low-energy theory for the partial ferrimagnetic phase is presented.
}

\kword
{mixed diamond chain, distortion, frustration, Haldane phase, spontaneous translational symmetry breakdown, partial ferrimagnetism
}

\begin{document}
\sloppy
\maketitle
\section{Introduction}
Quantum magnetism in frustrated spin systems is a rapidly developing field of condensed matter physics.\cite
{hfm2008,diep}
At first glance, one would expect %It may be expected
 that geometrical frustration enhances quantum fluctuation and drives an ordered state into a disordered
state.
However, recent progress in this field of physics has shown that this simple intuition is not always valid and that geometrical
frustration induces a variety of exotic quantum phenomena, which are not easily predicted.
Under an appropriate condition, it even stabilizes an unexpected magnetic long range order such as the frustration-induced
ferrimagnetic and spin nematic orders.

%Quantum magnetism in frustrated spin systems is a rapidly developing field of condensed matter physics.\cite{hfm2008,diep} 
%Usually, it is natural to expect that geometrical frustration enhances  quantum fluctuation and drives an ordered state  into a disordered state.  Recent progress in this field of physics, however,   has shown that this intuition is not always valid and that geometrical frustration can induce a variety of exotic quantum phenomena unpredictable by the simple-minded intuition.  Under an appropriate condition, it can even stabilize an unexpected    magnetic long range order such as the frustration-induced ferrimagnetic order and spin nematic order. 

To understand magnetism under the interplay of geometrical frustration and quantum fluctuation, 
it is desirable to begin with typical  spin models with exact solutions. 
Among them, there exist a class of models whose ground states are exactly written down as  spin cluster solid (SCS) states because of frustration.  
A SCS state is a tensor product state of exact local eigenstates of cluster spins. 
Well-known examples are the Majumdar-Ghosh model\cite{mg} whose ground state is a prototype of spontaneously dimerized phases in one-dimensional frustrated magnets\cite{hase} and  the Shastry-Sutherland model\cite{shs}  
which corresponds to the material SrCu$_2$(BO$_3$)$_2$.\cite{kage1,kage2} In these models, the spin clusters are singlet dimers.

\begin{figure} %[tbh] %Fig_1
%\centerline{\includegraphics[width=4.5cm]{lattice_dia.eps}}
\centerline{\includegraphics[width=4.5cm]{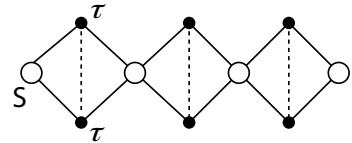}}
\caption{
Structure of the diamond chain. 
Spin magnitudes in a unit cell are indicated by $S$ and $\tau$; 
we denote the set of magnitudes by ($S$, $\tau$). 
The PDC is the case of $S = \tau$, while the MDC is the case of $S = 2\tau$ with an integer or half-odd integer $\tau$.}
\label{lattice_structure}
\end{figure}

The diamond chain is another frustrated spin chain with exact SCS ground states. 
The lattice structure is shown in Fig.~\ref{lattice_structure}. 
In a 
 unit cell, there are two kinds of nonequivalent lattice sites occupied by spins with magnitudes $S$ and $\tau$; 
we denote the set of magnitudes by ($S$, $\tau$). 
One of the authors and coworkers\cite{takano,Takano-K-S} introduced this lattice structure and generally investigated the case of ($S$, $S$), i.e., the pure diamond chain (PDC). 
Any PDC is shown to have at least one 
 exact SCS ground-state phase where each spin cluster has spin 0. 
Particularly, in the case of (1/2, 1/2), they determined the full phase diagram of the ground state by combining rigorous arguments with numerical calculations. 
After that, Niggemann et al.\cite{nig1,nig2} argued about 
a series of diamond chains with ($S$, 1/2). 
As for the special case of (1/2, 1/2), they reproduced 
the results of ref. \citen{Takano-K-S}. 

%====================================
%====================================

The mixed diamond chain (MDC) is defined as a diamond chain with ($S$, $S/2$) for the integer $S$.\cite{tsh} 
The special case of (1, 1/2) was first investigated 
by Niggemann et al.\cite{nig1,nig2} 
They considered it as one of the series of models with ($S$, 1/2). 
Recently, extensive investigation on the MDC has been carried out by the present authors.\cite{tsh,hts,htsalt} 
The MDC is of special interest among diamond chains, 
because only the MDC has the Haldane phase in the absence of 
 frustration, 
  so that we can observe the transition from the Haldane phase to a SCS phase induced by frustration. In contrast, diamond chains of  other types have  ferrimagnetic ground states  for weak frustration.

The features common to all types of diamond chains are their infinite number of local conservation laws and more than two different types of exact SCS ground states that are realized depending on the strength of frustration. 
For example, $S=1/2$ PDC has a nonmagnetic phase accompanied by spontaneous translational symmetry breakdown (STSB) and  a paramagnetic phase without STSB. 
This model also has a ferrimagnetic ground state in the less frustrated region.\cite{Takano-K-S} 
On the other hand, the MDC with spins 1 and $1/2$  has  3 different paramagnetic phases accompanied by STSB and one paramagnetic phase without STSB. 
This model  also has a nonmagnetic Haldane ground state in a less frustrated region.\cite{nig1, tsh} The SCS structures of the ground states are also reflected in characteristic thermal properties, as reported in ref. \citen{hts}. 

Modifications of the PDC and MDC have been examined by many authors. 
Among them, the spin 1/2 PDC with distortion has been thoroughly investigated by numerical methods.\cite{ottk,otk,sano} 
It is found that azurite, a natural mineral, consists of distorted PDCs with spin 1/2 and that the magnetic properties of this material have been experimentally studied in detail.\cite{kiku,ohta} 
Other materials have also been reported.\cite{izuoka,uedia} 
The diamond chain is  one of the simplest models compatible with the 4-spin cyclic interaction.  
The effects of this type of interaction on PDC have recently been investigated by Ivanov {\it et al.}\cite{dia4spin} 
The present authors also investigated the MDC with bond-alternating distortion and found an infinite series of ground states with STSB.\cite{htsalt} In addition, as reviewed in ref. \citen{htsalt}, 
the MDC is related to other important models of frustrated magnetism such as the dimer-plaquette model\cite{plaq,plaq2,koga1,koga2,plaq3,plaq4}, frustrated Heisenberg ladders\cite{frulad1}, hybrid diamond chains consisting of Heisenberg bonds and Ising bonds, \cite{str1,str2} and 
an Ising model on a hierarchical diamond lattice.\cite{fuku} Among them, the dimer-plaquette chain with ferromagnetic interplaquette interaction reduces to the MDC in the limit of strong interplaquette interaction.\cite{plaq4}

Thus far, in spite of the theoretical relevance of the MDC, no materials described by the MDC have been found. Nevertheless, synthesizing MDC materials is not an unrealistic expectation in view of the success of the synthesis of many low dimensional bimetallic magnetic compounds\cite{m-d}  and organic magnetic compounds.\cite{cb} 
In general, it is natural to expect that the lattice is possibly distorted in real MDC compounds as in azurite. From this viewpoint, it is important to present theoretical predictions on the ground state of distorted MDCs to widen the range of candidate materials of MDC and to raise the possibility of their synthesis. 
%=====================================
\begin{figure}[t] % Fig.2
%\centerline{\includegraphics[width=6cm]{mode.eps}}
\centerline{\includegraphics[width=6cm]{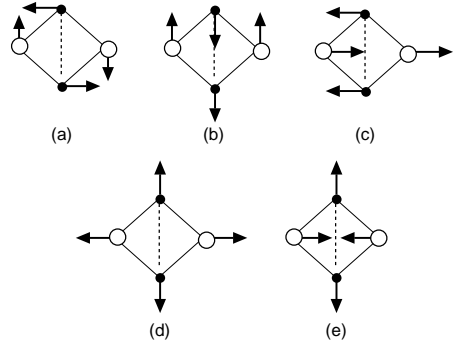}}
\caption{Displacement modes of a diamond unit.}
\label{mode}
\end{figure}
%=====================================

\begin{figure}[t] %Fig.3
%\centerline{\includegraphics[width=6cm]{lattice_all.eps}}
\centerline{\includegraphics[width=6cm]{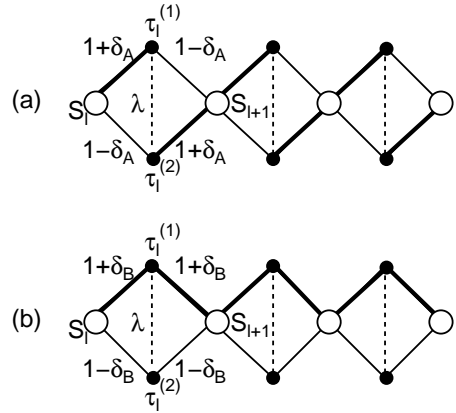}}
\caption{Structures of MDC 
 with {$S=1$ and $\tau^{(1)}=\tau^{(2)}=1/2$} with (a) type A and (b) type B distortions.}
\label{lattice}
\end{figure}

We begin by classifying the distortion patterns by the normal modes of each diamond unit. 
Excluding two translations and one rigid body rotation, we have 5 normal modes as depicted in Fig. \ref{mode} within the diamond plane. 
A distorted MDC may be  realized as a result of the collective softening of these normal modes. 
In particular, the distortion patterns in (a) and (b) break the local conservation laws that hold in the undistorted MDC. 
Hence, these distortions induce effective interactions between the cluster spins in the whole lattice, and may form novel exotic phases. 
We investigate these interesting cases in the present paper. 
In what follows, we name the distortion patterns in (a) and (b) 
as type A and type B, respectively. 
The MDCs with type A and type B distortions are depicted in Figs. \ref{lattice}(a) and \ref{lattice}(b), respectively. 
The distortion patterns in Figs. \ref{mode}(d) and \ref{mode}(e) do not change the geometry of the original undistorted MDC. 
The distortion pattern in Fig. \ref{mode}(c) is of  another interest, since it induces the bond alternation in the undistorted MDC 
without breaking the local conservation laws. 
This case has been investigated separately and published in a previous paper.\cite{htsalt}
%This case has been investigated in a separate paper.\cite{htsalt} 

%=====================================
%=====================================

This paper is organized as follows. 
In \S 2, the Hamiltonians for the MDCs with the type A and type B distortions are presented, and the structure of the ground states of the MDC without distortion is summarized. 
The ground-state phases for the MDC with the type A distortion are discussed in \S 3, and those for the MDC with the type B distortion are discussed in \S 4. 
The last  section is devoted to summary and discussion.

\section{Hamiltonian}
\label{section:ham}

The MDCs with the type A and type B distortions are described, respectively, by the following Hamiltonians: 
%------------------------------------------------------------
\begin{align}
{\cal H}_{\rm A} = &\sum_{l=1}^{N} 
\Bigl[ (1+\deltaa)\v{S}_{l}\v{\tau}^{(1)}_{l}
+ (1-\deltaa)\v{\tau}^{(1)}_{l}\v{S}_{l+1}
\nonumber\\
&+(1-\deltaa)\v{S}_{l}\v{\tau}^{(2)}_{l} 
+(1+\deltaa)\v{\tau}^{(2)}_{l}\v{S}_{l+1} 
+ \lambda\v{\tau}^{(1)}_{l}\v{\tau}^{(2)}_{l} \Bigr] , 
\label{hama}\\
{\cal H}_{\rm B} =&\sum_{l=1}^{N} 
\Bigl[ (1+\deltab)\v{S}_{l}\v{\tau}^{(1)}_{l}
+(1+\deltab)\v{\tau}^{(1)}_{l}\v{S}_{l+1}
\nonumber\\
&+(1-\deltab)\v{S}_{l}\v{\tau}^{(2)}_{l}
+(1-\deltab)\v{\tau}^{(2)}_{l}\v{S}_{l+1} 
+ \lambda\v{\tau}^{(1)}_{l}\v{\tau}^{(2)}_{l} \Bigr] , 
\label{hamb}
\end{align}
%------------------------------------------------------------
where $\v{S}_{l}$ is the spin-1 operator, and 
$\v{\tau}^{(1)}_{l}$ and $\v{\tau}^{(2)}_{l}$ 
are the spin-1/2 operators in the $l$th unit cell. 
The parameter $\deltaa$ ($\deltab$) represents the strength of type A (type B) distortion, and is taken to be nonnegative without spoiling generality. 
The number of unit cells is denoted by $N$, 
and then the total number of sites is $3N$. 
We will consider these systems in the large $N$ limit. 

For $\deltaa$ = 0 and $\deltab$ = 0, both eqs.~(\ref{hama}) and (\ref{hamb}) reduce to the undistorted MDC Hamiltonian, 
%------------------------------------------------------------
\begin{align}
{\cal H}_0 &=\sum_{l=1}^{N} \left[\v{S}_{l}\v{T}_{l}+\v{T}_{l}\v{S}_{l+1}+ \frac{\lambda}{2}\left(\v{T}^2_{l}-\frac{3}{2}\right)\right] 
\label{ham2}
\end{align}
%------------------------------------------------------------
with the composite spin operators 
$\v{T}_l\equiv\v{\tau}^{(1)}_{l}+\v{\tau}^{(2)}_{l}$. 
Before going into the analysis of the distorted MDC, we briefly summarize the ground-state properties of the Hamiltonian (\ref{ham2}) reported in ref. \citen{tsh} for convenience.
%%%%%
\begin{enumerate}
\item 
$\v{T}_l^2$ commutes with the Hamiltonian ${\cal H}_0$ for any $l$. 
Therefore, the composite spin magnitude $T_l$ 
 defined by $\v{T}_l^2 = T_l(T_l+1)$ is a good quantum number that takes the values 0 or 1. 
Hence, each energy eigenstate has a definite set of $\{T_l\}$, i.e. a sequence of 0's and 1's with length $N$. 
A pair of $\v{\tau}^{(1)}_{l}$ and $\v{\tau}^{(2)}_{l}$ with $T_l=0$ is called a dimer. 
 A cluster including $n$ successive $T_l=1$ pairs bounded by two $T_l=0$ pairs is called a cluster-$n$. 
The cluster-$n$ is equivalent to an antiferromagnetic spin-1 Heisenberg chain of length $2n+1$ with open boundary condition. Since a cluster-$n$ is decomposed into a sublattice consisting of $n+1$ sites with $\v{S}_l$'s and that consisting of $n$ sites with $\v{T}_l$'s, the ground states of a cluster-$n$ are spin triplet states with total spin unity on the basis of the Lieb-Mattis theorem.\cite{lm,kene} This implies that each cluster-$n$ carries a spin-1 in its ground state.

\item There appear 5 distinct ground-state phases called dimer-cluster-$n$ (DC$n$) phases with $n=0,1,2,3$, and $\infty$. 
The DC$n$ state %with finite $n$ 
is an alternating array of dimers and cluster-$n$'s. 
The phase boundary $\lambda_{\rm c}(n,n')$ between DC$n$ and DC$n'$ phases are
%------------------------------------------------------------
\begin{align}
&\lambda_{\rm c}(0,1) = 3, \nonumber\\
&\lambda_{\rm c}(1,2) \simeq 2.660425045542, \nonumber\\
&\lambda_{\rm c}(2,3) \simeq 2.58274585704, \nonumber\\ 
&\lambda_{\rm c}(3,\infty) \simeq 2.5773403291, 
\label{lambdac}
\end{align}
%------------------------------------------------------------
where $\lambda_{\rm c}(0,1)$ is obtained analytically and other values %the others
 are calculated numerically.
\item In the DC$\infty$ ground state realized for $\lambda < \lambda_{\rm c}(3,\infty) $,  
$T_l=1$ for all $l$. This state 
is not accompanied by STSB and 
is equivalent to the Haldane state of an antiferromagnetic spin-1 Heisenberg chain with infinite length. 
\item Each of the DC$n$ states with $0 \leq n \leq 3$ realized for  $\lambda > \lambda_{\rm c}(3,\infty)$ 
is a uniform array of cluster-$n$'s  with a common value of $n$ and dimers in between. 
In the DC$n$ phase with $1\leq n \leq 3$, $(n+1)$-fold STSB takes place.  In the DC$0$ phase, no translational symmetry is broken.
\end{enumerate}
%%%%%}
In what follows, we numerically examine various aspects of the type A and type B distortion effects on the MDC. 
Because the DC3 phase is only realized 
 within a very narrow interval of $\lambda$, it is difficult to analyze the effect of distortion numerically in this 
 phase. Hence, we do not consider  the DC3 phase in the following numerical analysis.

\section{Ground-State Properties of the MDC with Type A Distortion}

\subsection{Weak distortion regime}

%=====================================
\begin{figure} [t] %[!h] %Fig.4
%\centerline{\includegraphics[width=7cm]{effcp1.eps}}
%\centerline{\includegraphics[width=7cm]{effcp2end.eps}}
%\centerline{\includegraphics[width=7cm]{effcp3end.eps}}
\centerline{\includegraphics[width=6.5cm]{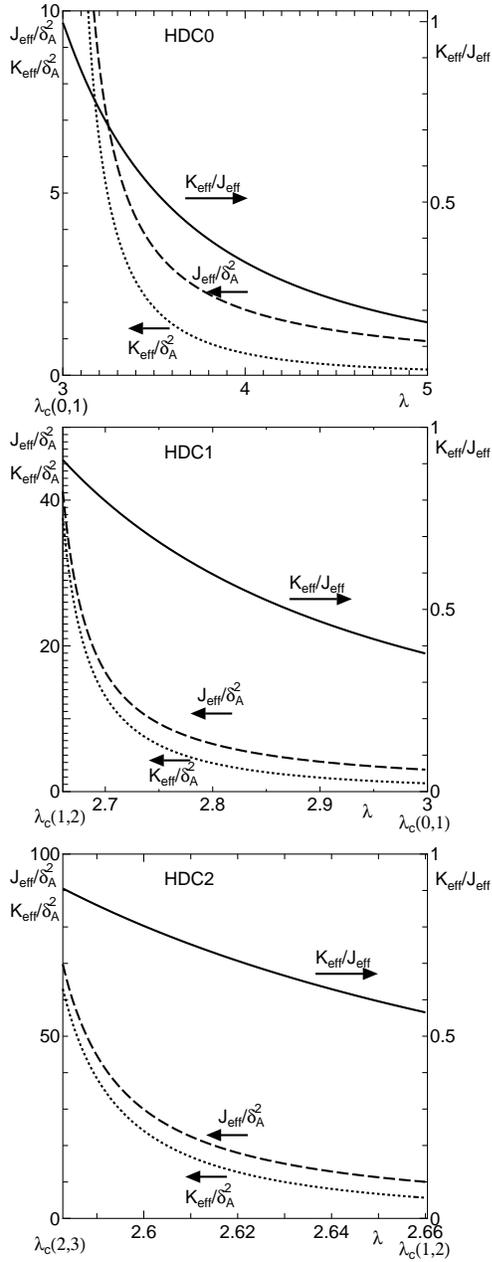}}
\caption{Effective bilinear interaction ($J_{\rm eff}$) and biquadratic interaction ($K_{\rm eff}$) between spin clusters 
for small $\delta_A$ in  HDC0, HDC1, and  HDC2 phases from top to bottom. The ratio $K_{\rm eff}/J_{\rm eff}$ is also shown.}
\label{jeff}
\end{figure}
%=====================================

We now inspect the nature of the effective interaction between two cluster-$n$'s  
 separated by a dimer consisting of $\v{\tau}_l^{(1)}$ and $\v{\tau}_l^{(2)}$ in the presence of the weak
  type A distortion. 
For $\deltaa> 0$, ${\v{S}}_l$ (${\v{S}}_{l+1}$) tends to be antiparallel to $\v{\tau}_l^{(1)}$  ($\v{\tau}_l^{(2)}$) rather than to $\v{\tau}_l^{(2)}$  ($\v{\tau}_l^{(1)}$), as is known from Fig.~\ref{lattice}(a). 
The spins $\v{\tau}_l^{(1)}$ and $\v{\tau}_l^{(2)}$ are antiparallel to each other because they form a singlet dimer. Therefore, ${\v{S}}_l$ and ${\v{S}}_{l+1}$ tend to be antiparallel to each other. In each cluster-$n$, the number of spins $\v{S}_l$'s is larger than the number of composite spins $\v{T}_l$'s by one. Hence, from the Lieb-Mattis theorem,\cite{lm} 
 the total spin of the ground state of the cluster-$n$ points to the same direction as the ${\v{S}}_l$'s belonging to that cluster-$n$. Therefore, the total spins of the cluster-$n$'s on both sides of the 
dimer also tend to be antiparallel to each other. 
Thus, the effective coupling between the spins of neighboring cluster-$n$'s is antiferromagnetic. 
This physical argument will be numerically ensured below.

In general, the interaction between two spins with a magnitude of  1 is the sum of 
bilinear and biquadratic terms. 
Therefore, the effective Hamiltonian for cluster-$n$'s in the phase that continues to the DC$n$ phase in the limit of $\deltaa\rightarrow 0$ is written as 
%------------------------------------------------------------
\begin{align}
{\cal H}^{\rm eff}&=\sum_{i=1}^{N_{\rm c}} {\cal H}^{\rm eff}(i,i+1), \\
{\cal H}^{\rm eff}(i,i+1)&=J_{\rm eff}\hat{\v{S}}_i\hat{\v{S}}_{i+1}+K_{\rm eff}\left(\hat{\v{S}}_i\hat{\v{S}}_{i+1}\right)^2, 
\end{align}
%------------------------------------------------------------
where $\hat{\v{S}}_i$ is the total 
 spin of the $i$-th cluster-$n$ with a magnitude of  1, $N_{\rm c}$ is the total number of cluster-$n$'s, and $J_{\rm eff}$ and $K_{\rm eff}$ are  effective coupling constants. From symmetry consideration, the signs of $\deltaa$ does not affect the sign of the effective coupling constants. Hence, 
  these coupling constants are 
  of the order of $\deltaaa$ for small $\deltaa$. 
We numerically calculated the ground-state energy of a pair of cluster-$n$'s with  total spin  $S_{\rm tot}$, and compared it with the corresponding eigenvalues of ${\cal H}^{\rm eff}(i,i+1)$. 
Then we confirmed 
 that $J_{\rm eff}/\deltaaa$ and $K_{\rm eff}/\deltaaa$ are almost independent of $\deltaa$ typically for $\deltaa < 0.002$. 
The constant values of $J_{\rm eff}/\deltaaa$ and $K_{\rm eff}/\deltaaa$ are shown in Fig. \ref{jeff} for three phases ($n$=0, 1, and 2), which will be explained below. 
 
Because the effective coupling constants satisfy  $0 < \Keff/\Jeff <1$, the ground state is 
the Haldane state for small $\deltaa$.\cite{ft}
In the Haldane state, each spin-1 degree of freedom is carried by a cluster-$n$ rather than by a single spin. 
We call the state 
the Haldane DC$n$ (HDC$n$) state. %s. 
In the HDC$n$ state with $n \geq 1$, the $(n+1)$-fold translational symmetry is spontaneously broken 
 unlike the conventional Haldane state without STSB.  
Both the 
HDC0 state  for $\lambda > \lambda_{\rm c}(0,1)$ and the HDC$\infty$ state for $\lambda < \lambda_{\rm c}(3,\infty)$ are the Haldane states without STSB. 
In particular, the HDC$\infty$ state continues from the  Haldane {state (DC$\infty$ state)} of the undistorted MDC mentioned 
in \S \ref{section:ham}.\cite{tsh} 

\subsection{Connection to the strong distortion regime}

\begin{figure} %Fig.5
%\centerline{\includegraphics[width=8cm]{hdcssqb.eps}}
\centerline{\includegraphics[width=8cm]{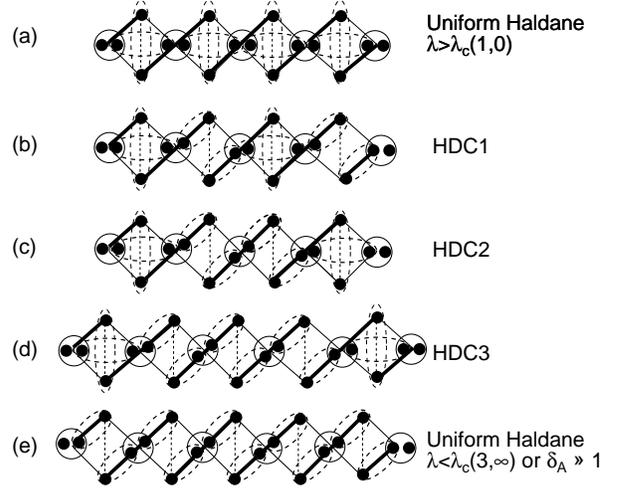}}
\caption{
Valence bond structures of the ground states of all phases for the MDC with type A distortion. 
A small filled circle represents a spin with a magnitude of  1/2. 
An original spin with a magnitude of  1 is represented by two decomposed 1/2 spins in an open circle indicating the symmetrization.  
A valence bond is represented by a dashed oval.} 
\label{valence}
\end{figure}

In the strong distortion regime of $\deltaa \rightarrow 1$ and small $\lambda$, 
the three spins $\v{\tau}^{(2)}_{l-1}$, $\v{S}_{l}$, and $\v{\tau}^{(1)}_{l}$ form a singlet cluster. 
Hence, the ground state is a state with spin gap and without STSB. 
This nature is common to the HDC0 and HDC$\infty$ phases in \S 3.1. 
Furthermore, the HDC$\infty$ state is transformed into
 the HDC0 state only by rearranging two valence bonds within each diamond unit, as shown in Figs.~\ref{valence}(a) and \ref{valence}(e). 
Therefore, the strong distortion, 
HDC0, and HDC$\infty$ regimes 
are considered to be different parts of a single phase. 
The continuity of the three regimes will be confirmed by the numerical analysis discussed in \S 3.3. 
In what follows, we call this phase 
the uniform Haldane (UH) phase as a whole.
%=====================================
%=====================================

\subsection{Numerical phase diagram}
%=====================================
\begin{figure} %Fig.6
%\centerline{\includegraphics[width=7cm]{lm280prof.eps}}
%\centerline{\includegraphics[width=7cm]{lm262prof.eps}}
\centerline{\includegraphics[width=7cm]{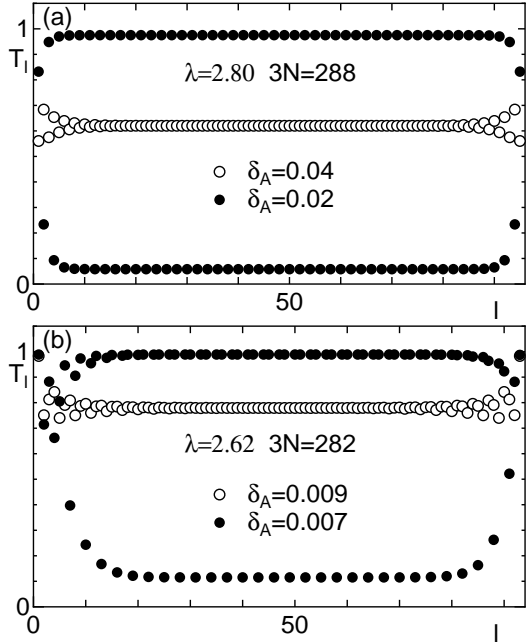}}
\caption{Profiles of $T_l$ for (a) $\lambda=2.8$  and (b) $\lambda=2.62$.}
\label{profile}
\end{figure}
%=====================================
%=====================================
\begin{figure} %Fig.7
%\centerline{\includegraphics[width=7cm]{lmd285dltndep.eps}}
%\centerline{\includegraphics[width=7cm]{lmd262dltndep.eps}}
\centerline{\includegraphics[width=7cm]{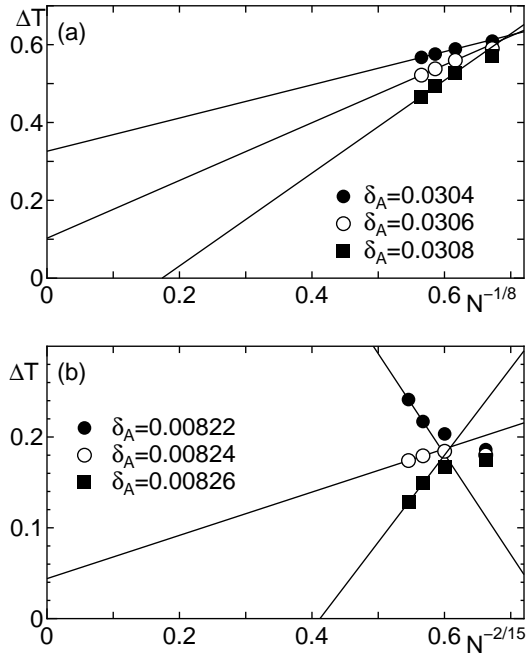}}
\caption{System size dependences of $\Delta T$ at (a) $\lambda = 2.85$ and (b) $\lambda = 2.62$. The data are plotted against $N^{-\beta/\nu}$ where $\beta$ and $\nu$ are the critical exponents of the order parameter and correlation length, respectively, for the 2-dimensional (a) Ising and (b) 3-clock model. }
\label{ndep}
\end{figure}
%=====================================
%=====================================
\begin{figure} %Fig.8
%\centerline{\includegraphics[width=7cm]{critical_dms.eps}}
\centerline{\includegraphics[width=7cm]{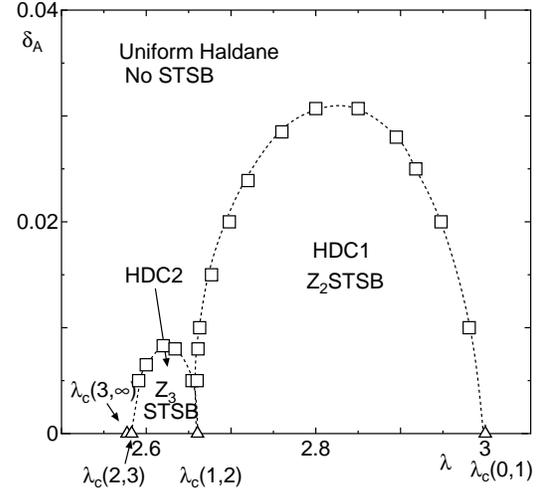}}
\caption{Phase diagram of the MDC with type A distortion. The triangles indicate the position of the phase boundary for $\deltaa=0$.}
\label{phase_stag}
\end{figure}
%=====================================
\begin{figure} %Fig.9
%\centerline{\includegraphics[width=7cm]{fscsa.eps}}
\centerline{\includegraphics[width=7cm]{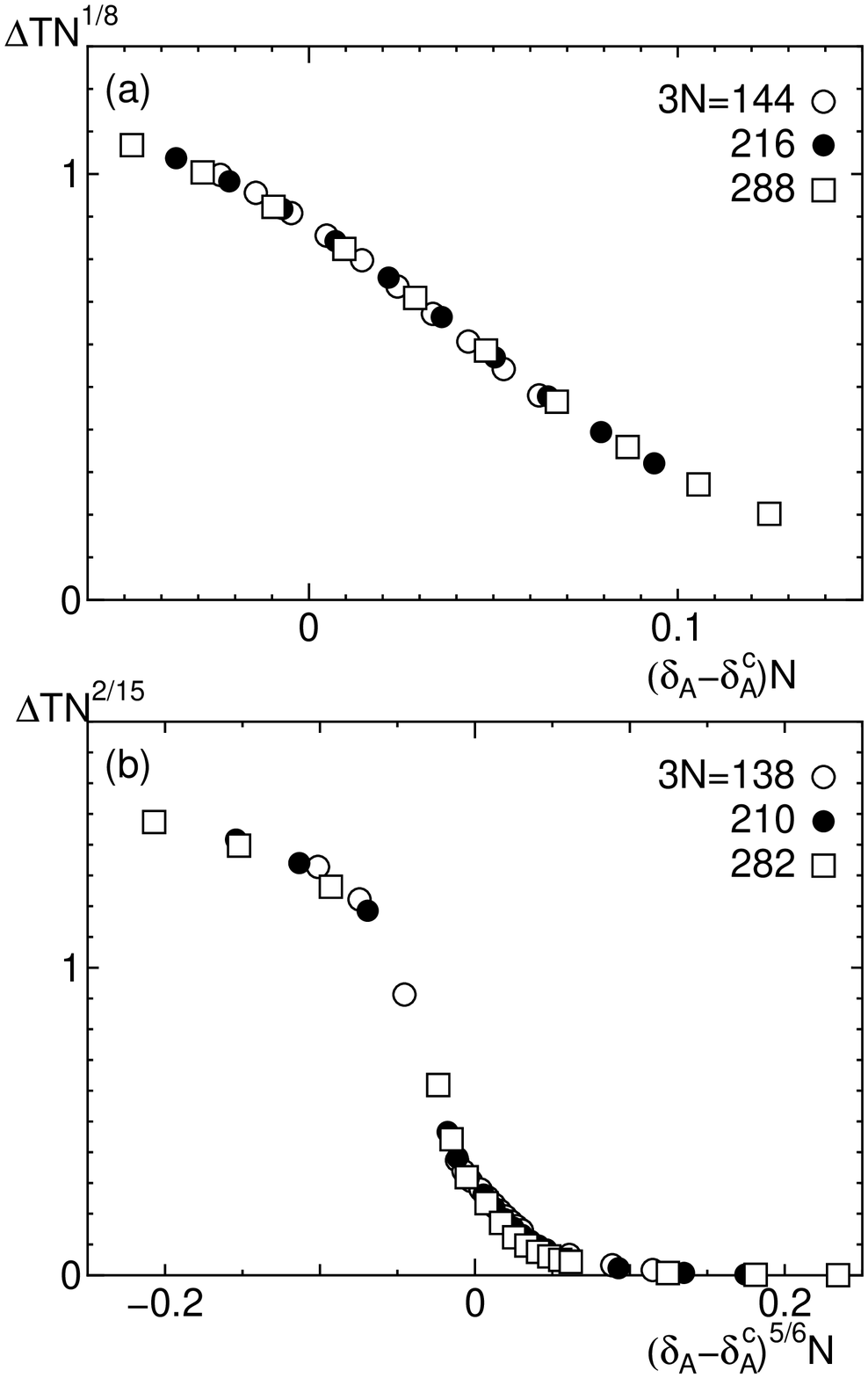}}
\caption{Finite-size scaling plot of $\Delta T$ around the critical points.  (a) Plot around the HDC1-UH phase boundary at $\lambda = 2.85$. The Ising critical exponents $\nu=1$ and $\beta=1/8$ are assumed. The critical point is set at $\delta_{\rm A}^{\rm c}=0.0307$. (b) Plot around the HDC2-UH phase boundary at  $\lambda = 2.62$. The 3-state Potts critical exponents $\nu=5/6$ and $\beta=1/9$\cite{wu} are assumed. The critical point is  set at $\delta_{\rm A}^{\rm c}=0.008248$.}
\label{z23}
\end{figure}

Under the periodic boundary condition, even in the parameter region where STSB takes place,  the ground state of a finite chain is a superposition of the symmetry-broken states, and the translational symmetry is recovered. Under the open boundary condition, however, one of the symmetry broken states is selected by the boundary effect. Therefore, we employ the DMRG calculation with the open boundary condition to determine the phase diagram for finite $\deltaa$. The DMRG calculation is carried out using the finite-size algorithm up to 288 sites keeping 200 states in subsystems. We calculate the ground-state expectation values  $\aver{\v{T}_l^2}$  and define the effective spin magnitude $T_l$ on the $l$-th diagonal bond by $T_l(T_l+1)=\aver{\v{T}_l^2}$. A typical $l$ dependence of $T_l$ is shown in Fig. \ref{profile} in each phase. With the increase in $\deltaa$, the translational symmetry is recovered as expected.  For finite $\deltaa$, the ground-state phase is identified from the periodicity in the oscillation of $T_l$. In the HDC$n$ phase, the values of $T_l$ follow the sequence 
%------------------------------------------------------------
\begin{align}
...T_{\rm S} \, \underbrace{T_{\rm L} \cdots T_{\rm L}}_{n} \, T_{\rm S}, \underbrace{T_{\rm L} \cdots T_{\rm L}}_{n} \,.., \ (T_{\rm L} > T_{\rm S}).
\end{align}
%------------------------------------------------------------
Thus, we define the order parameter of the HDC$n$ phase by $\Delta T=T_{\rm L}-T_{\rm S}$. In DMRG, $\Delta T$ is measured at the sites closest to the center of the chain. 

The valence bond structures for the HDC$n$ phases as well as the UH phase are shown in Fig. \ref{valence}. 
We see the translational invariance of period $n+1$ in the HDC$n$ ground state in contrast to the period-1 invariance in the UH ground state. 
Hence, the $Z_{n+1}$ STSB  takes place at the HDC$n$-UH phase boundary.  
We expect that this transition belongs to the 2-dimensional $(n+1)$-clock model universality class. 
The system size dependence of $\Delta T$ for $\lambda=2.85$ is shown in Fig. \ref{ndep}(a) around the HDC1-UH phase boundary. Here, the data are plotted against $N^{-\beta/\nu}$ with the order parameter critical exponent $\beta=1/8$ and the correlation length critical exponent $\nu=1$ for 
the two-dimensional Ising  universality class. 
This shows that the critical value of $\deltaa$ lies between 0.0304 and 0.0308. A similar plot is shown in Fig. \ref{ndep}(b) for $\lambda=2.62$ around the HDC2-UH phase boundary, assuming the critical exponents of two-dimensional 3-clock model  (equivalently 3-state Potts model\cite{wu}) with $\beta=1/9$ and $\nu=5/6$. This shows that the critical value of $\deltaa$ lies between 0.00822 and 0.00826. The critical points at other values of $\lambda$ are determined similarly.  
 The results are shown in the phase diagram of Fig. \ref{phase_stag}. The error bars are within the size of the symbols.

To confirm the consistency of the universality class, the finite size scaling 
 plot for the order parameter $\Delta T$ is carried out. According to the scaling hypothesis, the $\deltaa$ dependence of the order parameter $\Delta T$ of the finite size systems near the critical point should obey the finite size scaling law\cite{barbar}
\begin{align}
\Delta T N^{\beta/\nu}
=f(N(\deltaa-\delta_{\rm A}^{\rm c})^{\nu}),
\end{align}
in terms of the scaled variables  
$\Delta T N^{\beta/\nu}$ and 
$N(\deltaa-\delta_{\rm A}^{\rm c})^{\nu}$ and the scaling function $f(x)$.  
In Figs. \ref{z23}(a) and \ref{z23}(b), 
$\Delta T N^{\beta/\nu}$ 
is plotted against $N(\deltaa-\delta_{\rm A}^{\rm c})^{\nu}$ 
 around the HDC1-UH and HDC2-UH phase boundaries assuming the Ising and 3-clock universality classes, respectively. 
The critical points $\delta_{\rm A}^{\rm c}$ = 0.0307 (Fig.~\ref{z23}(a)) and 0.008248 (Fig.~\ref{z23}(b)) 
are chosen so that all data fall on a single universal scaling curve as well as possible. These plots are consistent with the expected universality class.

\begin{figure} %Fig.10
%\centerline{\includegraphics[width=8cm]{bilbiq.eps}}
\centerline{\includegraphics[width=7cm]{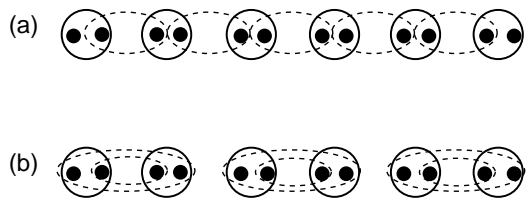}}
\caption{
Valence bond structures of the ground states of spin-1 bilinear-biquadratic chain in the (a) Haldane phase and (b) dimer phase. The spins with a magnitude of  unity represented by open circles are decomposed into two spin-1/2 degrees of freedom represented by small filled circles. The valence bonds are represented by dashed ovals. The spins belonging to  disconnected clusters in the dimer phase are connected by the valence bonds in the Haldane phase.
}
\label{bilbiq}
\end{figure}

The  critical behavior at the HDC$1$-UH transition in our model should be compared with that of the $S=1$ bilinear-biquadratic chain at the Takhtajan-Babujian point\cite{takh,babu}. 
 Both transitions are accompanied by $Z_2$-STSB which contributes to the conformal charge by 1/2. 
For the HDC$1$-UH transition in our model, the rearrangement of
valence bonds take place only within each diamond unit, as shown in
 Fig.~\ref{valence}(a) and \ref{valence}(b). 
In contrast, in the $S=1$ bilinear-biquadratic chain, the  spins  belonging to  disconnected clusters in the dimer phase are connected by the valence bonds in the Haldane phase, as  shown in Fig. \ref{bilbiq}. %In the latter  case
  Apart from $Z_2$-STSB, this is similar to the Gaussian criticality of the Haldane-dimer transition in the spin-1 alternating bond Heisenberg chain that contribute to the conformal charge by 1.\cite{ia,yt,kn}
Therefore,  the $S=1$ bilinear-biquadratic chain at the Takhtajan-Babujian point is described by the conformal field theory  with $c=1/2+1=3/2$, while the HDC$1$-UH transition in our model is described by the $c=1/2$ Ising conformal field theory. 

\section{Ground-State Properties of the MDC with Type B Distortion}

\begin{figure} %Fig.11
%\centerline{\includegraphics[width=6.5cm]{ferricrn.eps}}
\centerline{\includegraphics[width=7cm]{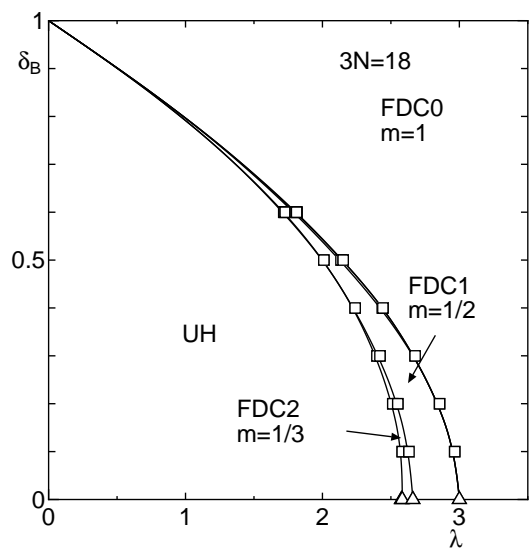}}
\caption{Phase diagram of the MDC with type B distortion with $3N=18$. The triangles indicate the position of the phase boundaries $\lambda_{\rm c}(n,n+1)$ for $\deltab=0$.\cite{tsh}}
\label{phase_ferri}
\end{figure}

In the case of type B distortion, the effective interaction between the spins of two cluster-$n$'s separated by the  
dimer consisting of $\v{\tau}_l^{(1)}$ and $\v{\tau}_l^{(2)}$ is ferromagnetic, because both $\v{S}_l$ and $\v{S}_{l+1}$ tend to be antiparallel to $\v{\tau}_l^{(1)}$.  
Therefore, we expect the ferrimagnetic ground state with spontaneous magnetization quantized as $m=$  
$1/(n+1)$ per unit cell for small $\deltab$ in the range $\lambda_{\rm c}(n,n+1) < \lambda < \lambda_{\rm c}(n-1,n)$. We call this phase a ferrimagnetic DC$n$ phase (FDC$n$ phase). 
In contrast, the ground state for $\lambda <\lambda_{\rm c}(3,\infty)$ will remain in the Haldane phase, since a nonmagnetic gapped phase is generally robust against a weak distortion. 
For finite $\deltab$, we determined the ground-state phase diagram  
by the numerical diagonalization for the system size $3N=18$, as shown in Fig.~\ref{phase_ferri}. 
Among system sizes tractable by  numerical diagonalization, only this size of $3N=18$ is compatible with all the ground-state structures with $n=0, 1$, and 2. 
As expected, the FDC$n$ phases with $m=1/(n+1)$ are found for these values of $n$.

%=====================================
%=====================================
%=====================================
\begin{figure} %Fig.12
%\centerline{\includegraphics[width=7cm]{magalmono.eps}}
\centerline{\includegraphics[width=7cm]{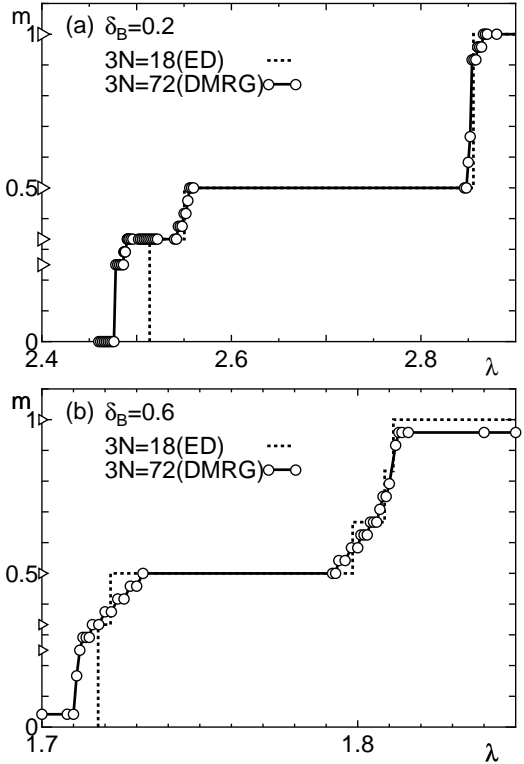}}
\caption{Spontaneous magnetization for (a) $\deltab=0.2$ and (b) $\deltab=0.6$. The triangles on the vertical axes 
indicate the values of the spontaneous magnetization $m=1/(n+1)$ 
in the FDC$n$ phases.}
\label{mag}
\end{figure}
%=====================================

By inspecting numerical data for the $3N=18$ system, 
we also find other narrow ferrimagnetic phases between the FDC$n$ and FDC$(n+1)$ phases with $n=$ 0, 1, and 2, although they are too narrow to be shown in Fig.~\ref{phase_ferri}. 
In order to investigate these phases in detail, we employ the DMRG calculation for 
$3N=72$ keeping 120 states in each subsystem.  
Typical examples of the $\lambda$ dependence of spontaneous  magnetization are shown in Fig. \ref{mag}(a) for $\deltab=0.2$ and Fig. \ref{mag}(b) for $\deltab=0.6$.  Between the FDC$n$ and FDC$(n+1)$ phases with $n=0,1,2$, we find the partial ferrimagnetic phase in which the spontaneous magnetization varies continuously with $\lambda$. The ferrimagnetic phase of this kind has been found in various frustrated one-dimensional quantum spin systems.\cite{ss,bkk,ir,ym,kh,htsdec,filho}  In contrast, between the nonmagnetic phase and the FDC$3$ phase, we find no partial ferrimagnetic phase for small $\deltab$. 

This can be understood as follows: At $\lambda=\lambda_{\rm c}(n,n+1)$, the cluster-$n$ and cluster-$(n+1)$ can coexist. As stated above, 
 it is physically evident that the effective magnetic interaction between the clusters is ferromagnetic. Therefore, we can restrict the states of each cluster to the  maximally polarized ground state with $\hat{S}^z_i=1$. Hence, the ground state of the whole chain is described by specifying the arrangement of  cluster-$n$'s and cluster-$(n+1)$'s. We map the two possible values of the length of $i$-th cluster, $n_i=n$ and $n_i=n+1$, to two possible values of the spin-1/2 pseudospin, $\sigma^z_i=1/2$ and $\sigma^z_i=-1/2$, respectively. Then, the total magnetization $M$ is equal to the number of clusters $N_{\rm c}$. The total number of  unit cells, $N$, is related to the pseudospins $\sigma^z_i$ as
%------------------------------------------------------------
\begin{align}
N=\sum_{i=1}^{N_{\rm c}} \left(n+1+\frac{1}{2}-\sigma^z_i\right)
=N_{\rm c}\left(n+\frac{3}{2}\right)-\sum_{i=1}^{N_{\rm c}} \sigma^z_i.
\label{eq:length}
\end{align}
%------------------------------------------------------------
Therefore, the ground-state magnetization per unit cell $m$ is given by
%------------------------------------------------------------
\begin{align}
m=\frac{N_{\rm c}}{\aver{N}}
=\frac{1}{n+\frac{3}{2}-\sigma} 
\end{align}
%------------------------------------------------------------
with $\sigma\equiv \sum_{i=1}^{N_{\rm c}} \aver{\sigma^z_i} /N_{\rm c}$. The bracket  
$\aver{\cdots}$ represents the ground-state expectation value. 
In the presence of $\deltab$, the length of neighboring clusters can exchange through a second order process 
 in $\deltab$. This corresponds to the spin exchange in terms of pseudospins. 
In this case, the interaction between the pseudospins is approximated by 
 the  spin-$1/2$ XXZ Hamiltonian
%------------------------------------------------------------
\begin{align}
\H_{\rm XXZ}&=
\sum_{i=1}^{N_{\rm c}}\H_{\rm XXZ}(i,i+1) , \\
\H_{\rm XXZ}(i,i+1)&=J_{z}^{\rm eff}\sigma^z_i\sigma^z_{i+1}+J_{\perp}^{\rm eff}(\sigma^x_i\sigma^x_{i+1}+\sigma^y_i\sigma^y_{i+1})\label{hameff0}
\end{align}
%------------------------------------------------------------
up to the second order in $\deltab$. Here, further neighbor interactions are neglected. We estimate 
 the effective exchange constants by comparing the energy spectrum of the pair Hamiltonian $\H_{\rm XXZ}(i,i+1)$ with that of the corresponding pair of clusters as follows:
\begin{enumerate}
\item $\lambda=\lambda_{\rm c}(0, 1)$
%------------------------------------------------------------
\begin{align}
J_z^{\rm eff}&\simeq -0.039\deltab^2, \nonumber\\
J_{\perp}^{\rm eff}&\simeq 0.087\deltab^2.\label{j01}
\end{align}
%------------------------------------------------------------
\item $\lambda=\lambda_{\rm c}(1, 2)$
%------------------------------------------------------------
\begin{align}
J_z^{\rm eff}&\simeq -0.0082\deltab^2, \nonumber\\
J_{\perp}^{\rm eff}&\simeq 0.069\deltab^2.\label{j12}
\end{align}
%------------------------------------------------------------
\item $\lambda=\lambda_{\rm c}(2, 3)$
%------------------------------------------------------------
\begin{align}
J_z^{\rm eff}&\simeq -0.0029\deltab^2, \nonumber\\
J_{\perp}^{\rm eff}&\simeq 0.018\deltab^2.\label{j23}
\end{align}
%------------------------------------------------------------
\end{enumerate}
{The  details of the calculation are explained in Appendix.}

In all cases (i) $\sim$ (iii), we find that the effective coupling constants satisfy the inequality $-|J_{\perp}^{\rm eff}| < J_{z}^{\rm eff} \leq |J_{\perp}^{\rm eff}|$. As is well known, the ground state of the spin-1/2 XXZ chain in this parameter regime is nonmagnetic and gapless in the absence of a magnetic field. 
Roughly speaking,  $\Delta\lambda\equiv\lambda-\lambda_{\rm c}(n,n+1)$ corresponds to 
the effective magnetic field $h_{\rm eff}$ conjugate to 
the total pseudospin $\sum_i \sigma^z_i$, 
because the increase in $\lambda$ favors  cluster-$n$ over cluster-$(n+1)$; however,  
this correspondence should not be taken literally. 
A more precise argument is {also} given in Appendix. 
When $\Delta\lambda$ 
takes a large negative value,  
the pseudospins are fully polarized downward to give $\aver{\sigma^z_l}=-1/2$. This state corresponds to the  FDC$(n+1)$ state with $m=1/(n+2)$.  When $h_{\rm eff}$ reaches the critical value $h_{\rm c1}$  
$\equiv -(|J_{\perp}^{\rm eff}|+J_{z}^{\rm eff})$, the magnetization starts to increase continuously until all spins are  fully polarized upward at the  critical effective field $h_{\rm c2}\equiv |J_{\perp}^{\rm eff}|+J_{z}^{\rm eff}$. This corresponds to the FDC$n$ state with $m=1/(n+1)$. 

On the other hand, the magnetization jumps from 0 to $1/4$ at the phase boundary between the Haldane phase and the FDC3 phase for small $\deltab$. At this phase boundary, no finite size clusters 
coexist with cluster-$3$. Therefore, no pseudospin degrees of freedom can be defined. Consequently, no partial ferrimagnetic phase can be realized. 
In contrast, for 
larger values of $\deltab$,  
we numerically find a partial ferrimagnetic phase between the FDC3 and UH phases. This would be ascribed to the contribution of 
  other finite-length clusters with low lying energies which come into play through higher-order processes in $\deltab$.
 
\section{Summary and Discussion}

We introduced two types of distortion, type A and type B, into the MDC with spins 1 and 1/2, and investigated the ground-state phases. 
The phase diagrams are characteristic of the type A and type B distortions, respectively. 
For the type A distortion, the effective interaction between the cluster spins is antiferromagnetic with bilinear and biquadratic terms. The numerically estimated values of the effective couplings show that the DC$n$ ground states are transformed into the HDC$n$ ground states. 
 The order parameters characterizing the HDC$n$ phases are defined and the UH-HDC$n$ phase boundaries are determined using the DMRG data. From the valence bond structure of each phase, we expect that the UH-HDC$n$ phase transition belongs to  the universality class of the 2-dimensional $(n+1)$-clock model. The finite size scaling plot of the order parameter is consistent with this identification. 
For the type B distortion, the effective interaction between the cluster spins is ferromagnetic. In addition to the FDC$n$ phases with quantized spontaneous magnetization $m=1/(n+1)$, 
the partial ferrimagnetic phases are also found numerically between the FDC$n$ and FDC$(n+1)$ phases. A physical interpretation of the partial ferrimagnetic phase is given for small $\deltab$ by mapping onto an effective pseudospin-1/2 XXZ chain. 
 
Generally, the introduction of lattice distortion into a physical model increases the possibility that a corresponding material is realized. 
In the MDC, there are three types of distortion modes affecting the exchange interactions. 
Among them, the two types  
investigated in the present paper are of generic nature, because the local conservation laws that hold 
 in the undistorted MDC  
are broken. 
This suggests that the observation of the exotic phenomena predicted in the present paper is possible even if the corresponding material is not exactly described by the model Hamiltonians (\ref{hama}) and (\ref{hamb}). 
  
If a distorted MDC material is synthesized, the distortion may  
be controlled by, e.g., applying pressure. 
If the distortion is of type A, the Curie constant vanishes as the DC$n$ ground state turns into one of the HDC$n$ ground states. 
The magnetic susceptibility and magnetic specific heat will have an activation-type temperature dependence with activation energy proportional to the effective coupling between the cluster spins, which is of the order of  $\deltaa^2$.  
These HDC$n$ phases are not realized if the distortion $\deltaa$ exceeds $\sim 0.03$ even in the most robust case of $n=1$. 
In a real material, the STSB in the valence bond structure manifests itself as a magnetic superstructure. It is also possible that it is accompanied by a lattice superstructure of corresponding periodicity if the spin-lattice coupling is present. Therefore careful measurements of magnetic and lattice superstructures would help with the observation of  HDC$n$ phases with $1 \leq n \leq 3$. 

On the other hand, if the distortion of the material is of type B, the ground sate is ferrimagnetic. 
At low but finite temperatures, however, the spontaneous magnetization vanishes owing to the one-dimensionality. 
As a precursor of ordering at $T=0$, the low-temperature magnetic susceptibility should diverge as $T^{-2}$ with a coefficient proportional to the effective coupling$\sim \deltab^2$ between the cluster spins.\cite{yamamoto,yf,takahashi} 
This means that even a weak magnetic field of the order of $H \sim T^2/\deltab^2$ 
derives the finite-temperature magnetization up to the value of the ground-state spontaneous magnetization. 
This enables the experimental estimation of the spontaneous magnetization in real materials. The quantized ferrimagnetic behavior should be observed for wide ranges of the parameters $\lambda$ and $\deltab$ as shown in Fig. \ref{phase_ferri}, and should be easily observed if an appropriate material is synthesized. The partial ferrimagnetic phases are limited to narrow intervals of the parameters $\deltab$ and $\lambda$. Therefore, these can only be observed as a temperature-independent crossover between two quantized ferrimagnetic behaviors with careful exclusion of the thermal effect. 

We have demonstrated that various exotic ground states 
and phase transitions between them are realized in  
the distorted MDC with spins 1 and 1/2, 
which has a strong frustration. 
The physical pictures of these phenomena have become clear. 
This is made possible because the ground state of the {\it undistorted} MDC is known exactly. Therefore, we expect that our model may provide a means of understanding the similar exotic phenomena realized owing to the interplay of spin ordering, quantum fluctuation, and strong frustration in more general frustrated quantum chains on a firm ground. 
For example, partial ferrimagnetic phases are found in various  one-dimensional frustrated quantum spin models\cite{ss,bkk,ir,ym,kh,htsdec,filho}. However, some of them are only numerically confirmed  and no physical explanation has been given so far. We hope that the present study paves the way to the general understanding of these partial ferrimagnetic states. 

We thank J. Richter for drawing our attention to ref.  \citen{plaq4} and related works. The numerical diagonalization program is based on the package TITPACK ver.2 coded by H. Nishimori.  The numerical computation in this work has been carried out using the facilities of the Supercomputer Center, Institute for Solid State Physics, University of Tokyo and Supercomputing Division, Information Technology Center, University of Tokyo.  KH is  supported by a Grant-in-Aid for Scientific Research  on Priority Areas, "Novel States of Matter Induced by Frustration" (20048003) from the Ministry of Education, Culture, Sports, Science and Technology of Japan and a Grant-in-Aid for Scientific Research (C) (21540379) from the Japan Society for the Promotion of Science. KT and HS are  supported by a Fund for Project Research from Toyota Technological Institute. 

\appendix
\section{}

The Hamiltonian ${\cal H}_{\rm B}$ with the type B distortion is rewritten as 
%------------------------------------------------------------
\begin{align}
{\cal H}_{\rm B} = {\cal H}_0 +\delta {\cal H} , 
\end{align}
%------------------------------------------------------------
where  
%------------------------------------------------------------
\begin{align}
\delta{\cal H} = \deltab\sum_{l=1}^{N} 
\bigl( \v{S}_{l} + \v{S}_{l+1} \bigr) 
\bigl( \v{\tau}^{(1)}_{l} - \v{\tau}^{(2)}_{l} \bigr) . 
\label{hamp}
\end{align}
%------------------------------------------------------------
For small $\deltab$, the ground state around $\lambda=\lambda_{\rm c}(n,n+1)$ consists almost entirely of cluster-$n$'s and cluster-$(n+1)$'s. 
Hence, as a good approximation, we consider ${\cal H}_{\rm B}$ only in the restricted Hilbert space where each state involves no clusters except for cluster-$n$'s and cluster-$(n+1)$'s. 
Under the fixed cluster number $\Nc$ in this Hilbert space, 
${\cal H}_0$ is equivalent to the following effective Hamiltonian expressed in terms of pseudospin operators: 
%------------------------------------------------------------
\begin{align}
{\cal H}^0_{\rm eff}&=E_{\rm G}^0(n+1 ; \lambda)\sum_{i=1}^{\Nc} \left(\frac{1}{2}-\sigma^z_i\right)\nonumber\\
& +E_{\rm G}^0(n ; \lambda)\sum_{i=1}^{\Nc} \left(\frac{1}{2}+\sigma^z_i\right) , 
\end{align}
%------------------------------------------------------------
where $\sigma^z_i=1/2$ and $\sigma^z_i=-1/2$ correspond to $n_i=n$ and $n_i=n+1$, respectively. $E_{\rm G}^0(n ; \lambda)$ is the ground-state energy of a cluster-$n$ and a dimer in the absence of distortion, and is given by
%------------------------------------------------------------
\begin{align}
E_{\rm G}^0(n ; \lambda)&=\EHal(2n+1)+\frac{\lambda n}{4}-\frac{3\lambda }{4} , 
\end{align}
%------------------------------------------------------------
where $\EHal(2n+1)$ is the ground-state energy of the spin-1 antiferromagnetic Heisenberg chain with length $2n+1$.\cite{tsh}

The application of $\delta{\cal H}$ to the unperturbed ground state transforms one of the $T_l=0$ bonds to a $T_l=1$ bond or vice versa. 
Then the resulting  
states contain clusters with lengths less than $n$ or greater than $2n$. 
Since these states are outside the restricted Hilbert space, 
no  correction  to the ground-state energy  is present  within the first order in $\deltab$. Hence, the lowest-order correction is of the order of $\deltab^2$. 
Up to the second order in $\deltab$, the effective pseudospin Hamiltonian is given by 
%------------------------------------------------------------
\begin{align}
{\cal H}_{\rm eff}&=E_{\rm G}(n+1;\lambda,\deltab)\sum_{i=1}^{\Nc} \left(\frac{1}{2}-\sigma^z_i\right)\nonumber\\
& +E_{\rm G}(n ; \lambda,\deltab)\sum_{i=1}^{\Nc} \left(\frac{1}{2}+\sigma^z_i\right) +{\cal H}_{\rm XXZ},  
\end{align}
%------------------------------------------------------------
where 
$E_{\rm G}(n ; \lambda,\deltab)$
is the ground-state energy of a cluster-$n$ and a dimer including the second order correction in $\deltab$.  
This is simply expressed as 
%------------------------------------------------------------
\begin{align}
{\cal H}_{\rm eff} 
={\Nc}\bar{E}_{\rm G}+\Delta E_{\rm G}\sum_{i=1}^{\Nc} \sigma^z_i+{\cal H}_{\rm XXZ}, 
\end{align}
%------------------------------------------------------------
with 
%------------------------------------------------------------
\begin{align}
\bar{E}_{\rm G}&=\frac{1}{2}(E_G(n+1;\lambda,\deltab)+E_G(n ; \lambda,\deltab)) , \\
\Delta E_{\rm G}&=E_G(n ; \lambda,\deltab)-E_G(n+1;\lambda,\deltab).
\end{align}
%------------------------------------------------------------

The effective coupling constants $J_z^{\rm eff}$ and $J_{\perp}^{\rm eff}$ in ${\cal H}_{\rm XXZ}$ are also of the second order in $\deltab$.  
We determine  $J_z^{\rm eff}$ and $J_{\perp}^{\rm eff}$ so as to reproduce the low-lying energy spectrum of a pair of  cluster-$n$'s by that of two-pseudospin Hamiltonian 
\begin{align}
{\cal H}_{\rm eff}(i,i+1)
&=2\bar{E}_{\rm G}+\Delta E_{\rm G} (\sigma^z_i+\sigma^z_{i+1})+{\cal H}_{\rm XXZ}(i,i+1) 
\end{align}
In each of the subspaces $\sigma^z_i+\sigma^z_{i+1}=\pm 1$, $\sigma^z_i=\sigma^z_{i+1}=\sigma=\pm 1/2$. Therefore, the Hilbert space is one-dimensional and the eigenvalue of ${\cal H}_{\rm eff}(i,i+1)$ is simply $E_{\sigma\sigma}=2\bar{E}_{\rm G}+2\Delta E_{\rm G} \sigma + J_z^{\rm eff}/4$ with $\sigma=\pm 1/2$. In the subspace $\sigma^z_i+\sigma^z_{i+1}=0$, the Hilbert space is two-dimensional and the eigenvalues of ${\cal H}_{\rm eff}(i,i+1)$ are $E_{\pm}=2\bar{E}_{\rm G} - J_z^{\rm eff}/4 \pm J_{\perp}^{\rm eff}/2$. 

The original Hamiltonian of the cluster consisting of a cluster-$n$ and a cluster-$n'$ is the distorted diamond chain with length $n+n'$. 
\begin{align}
{\cal H}&(n+n')=\sum_{l=1}^{n+n'+1} 
\Bigl[ (1+\deltab)\v{S}_{l}\v{\tau}^{(1)}_{l}
+(1+\deltab)\v{\tau}^{(1)}_{l}\v{S}_{l+1}
\nonumber\\
&+(1-\deltab)\v{S}_{l}\v{\tau}^{(2)}_{l}
+(1-\deltab)\v{\tau}^{(2)}_{l}\v{S}_{l+1} 
+ \lambda\v{\tau}^{(1)}_{l}\v{\tau}^{(2)}_{l} \Bigr] , 
\label{hamnn}
\end{align}
We denote the $\alpha$-th eigenvalue of ${\cal H}(n+n')$ as $E(n+n';\alpha)$. 
Comparing the corresponding expression for the eigenvalues, we find
\begin{align}
E(2n;0)&=E_{\frac{1}{2},\frac{1}{2}}=2\bar{E}_{\rm G}+\Delta E_{\rm G} +\frac{J_z^{\rm eff}}{4}\\
E(2n+2;0)&=E_{-\frac{1}{2},-\frac{1}{2}}=2\bar{E}_{\rm G}-\Delta E_{\rm G} +\frac{J_z^{\rm eff}}{4}\\
E(2n+1;0)&=E_{-}=2\bar{E}_{\rm G} -\frac{J_z^{\rm eff}}{4} - \frac{J_{\perp}^{\rm eff}}{2} \\
E(2n+1;1)&=E_{+}=2\bar{E}_{\rm G} -\frac{J_z^{\rm eff}}{4} + \frac{J_{\perp}^{\rm eff}}{2} .
\end{align}
Solving these sets of equations, with respect to $J_z^{\rm eff}$ and $J_{\perp}^{\rm eff}$, we find
\begin{align}
J_{\perp}^{\rm eff} &=E(2n+1;1)-E(2n+1;0)\label{jeff1} , 
\\
J_z^{\rm eff} &=2[E(2n+2;0)+E(2n;0)
\nonumber\\
& \qquad -E(2n+1;1)-E(2n+1;0)] . 
\label{jeff2}
\end{align}
Note that the rhs's of (\ref{jeff1}) and  (\ref{jeff2}) vanish for $\deltab=0$. 
We numerically evaluated $E(2n;0)$, $E(2n+1;0)$, $E(2n+1;1)$, and $E(2n+2;0)$ at $\lambda=\lambda_{\rm c}(n,n+1)$ ($n = 0, 1, 2$) for small $\deltab$. 
Using these values in eqs.~(\ref{jeff1}) and  (\ref{jeff2}), we  determined $J_{\perp}^{\rm eff}$ and $J_z^{\rm eff}$ as eqs. (\ref{j01})-(\ref{j23}). 

\begin{figure} %Fig.A1
%\centerline{\includegraphics[width=7cm]{sigall.eps}}
\centerline{\includegraphics[width=7cm]{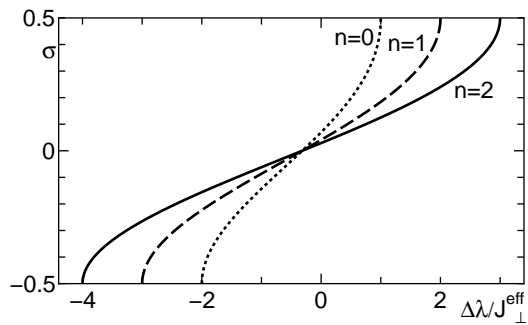}}
\caption{Relationship between $\sigma$ and $\Delta \lambda$ for $n=0, 1$, and 2.}
\label{effmag}
\end{figure}

For the whole MDC, the ground-state energy is written as 
%------------------------------------------------------------
\begin{align}
E_0&=\Nc \bar{E}_{\rm G}+\Nc\Delta E_G\sigma+\Nc\epsilon_{\rm XXZ}(\sigma) 
\end{align}
%------------------------------------------------------------
where  $\epsilon_{\rm XXZ}(\sigma)$ is the ground-state energy per site of a magnetized spin-1/2 XXZ chain with $\aver{\sigma_i^z}=\sigma$. The number of  unit cells, $N$, of the original MDC is given by the expectation value of eq.~(\ref{eq:length}) as $
N=N_{\rm c}(n+\frac{3}{2}-\sigma)
$. Therefore, we have
%------------------------------------------------------------
\begin{align}
E_0&=\frac{N}{n+\frac{3}{2}-\sigma}\left(\bar{E}_{\rm G}+\Delta E_G\sigma+\epsilon_{\rm XXZ}(\sigma) \right).
\end{align}
%------------------------------------------------------------
Minimizing this with respect to $\sigma$ with fixed $N$, we find 
%------------------------------------------------------------
\begin{align}
\Delta \lambda=\left(n+\frac{3}{2}-\sigma\right)\frac{\partial \epsilon_{\rm XXZ}(\sigma)}{\partial \sigma} +\epsilon_{\rm XXZ}(\sigma),
\end{align}
%------------------------------------------------------------
where $\Delta\lambda=\lambda-\lambda_{\rm c}(n,n+1; \deltab)$ and $\lambda_{\rm c}(n,n+1; \deltab)$ is defined by 
%------------------------------------------------------------
\begin{align}
(n+2)E_G(n;\lambda_{\rm c},\deltab)-(n+1)E_G(n+1;\lambda_{\rm c},\deltab)=0 . 
\end{align}
To simplify the calculation, we replace $\epsilon_{\rm XXZ}(\sigma)$  by the ground-state energy of the spin-1/2 XY chain 
$\epsilon_{\rm XY}=- (J_{\perp}^{\rm eff}/\pi) \cos\pi\sigma$, 
because  $|J^{\rm eff}_{\perp}|$ is substantially larger than $|J^{\rm eff}_{z}|$ in all cases. Then we find
%------------------------------------------------------------
\begin{align}
\frac{\Delta \lambda}{J_{\perp}^{\rm eff}}=\left(n+\frac{3}{2}-\sigma\right)\sin\pi\sigma-\frac{1}{\pi}\cos\pi\sigma.
\end{align}
%------------------------------------------------------------
This relation is plotted in Fig. \ref{effmag} for $n=0, 1$ and 2. It is clear that $\sigma$ continuously increases from $-1/2$ to $1/2$ with an increase in $\lambda$.  
%=====================================
%=====================================


\begin{thebibliography}{50}
\bibitem{diep} {\it Frustrated Spin Systems}, ,ed. H. T. Diep: (World Scientific, Singapore, 2005), Chaps. 5 and 6. 
\bibitem{hfm2008} {\it Proc. Int. Conf. on Highly Frustrated Magnetism (HFM2008)} J.  Phys.: Conf. Series {\bf 145} (2009). 
\bibitem{mg} C. K. Majumdar and D. K. Ghosh: J. Math. Phys. {\bf 10} (1969) 1399. 
\bibitem{hase} For examples of experimental materials, see M. Hase, H. Kuroe, K. Ozawa, O. Suzuki, H. Kitazawa, G. Kido, and T. Sekine: Phys. Rev. B {\bf 70} (2004) 104426.
\bibitem{shs} B. S. Shastry and B. Sutherland: Physica B+C {\bf 108} (1981) 1069.
\bibitem{kage1} H. Kageyama, K. Yoshimura, R. Stern, N.V. Mushnikov, K. Onizuka,
M. Kato, K. Kosuge, C.P. Slichter, T. Goto, and Y. Ueda: Phys. Rev. Lett. {\bf 82} (1999) 3168.
\bibitem{kage2} H. Kageyama, M. Nishi, N. Aso, K. Onizuka, T. Yosihama, K.
Nukui, K. Kodama, K. Kakurai, and Y. Ueda: Phys. Rev. Lett. {\bf 84} (2000) 5876. 
\bibitem{takano}
K. Takano: 
J. Phys. A: Math. Gen. {\bf 27} (1994) L269. 
\bibitem{Takano-K-S}
K. Takano, K. Kubo, and H. Sakamoto: 
J. Phys.: Condens. Matter {\bf 8} (1996) 6405. 
\bibitem{nig1} H. Niggemann, G. Uimin, and J. Zittartz: J. Phys.: Condens. Matter {\bf 9} (1997) 9031.
\bibitem{nig2} H. Niggemann, G. Uimin, and J. Zittartz: J. Phys.: Condens. Matter {\bf 10} (1998) 5217.
\bibitem{tsh}
K. Takano,  H. Suzuki, and K. Hida: 
Phys. Rev. B {\bf 80} (2009) 104410. 
\bibitem{hts}
K. Hida, K. Takano, and H. Suzuki: J. Phys. Soc. Jpn. {\bf 78} (2009) 084716
\bibitem{htsalt}
K. Hida, K. Takano, and H. Suzuki:  J. Phys. Soc. Jpn. {\bf 79} (2010) 044702.
\bibitem{ottk} K. Okamoto, T. Tonegawa, Y. Takahashi, and
M. Kaburagi: J. Phys.: Condens. Matter {\bf 11} (1999) 10485.
\bibitem{otk} K. Okamoto, T. Tonegawa, and M. Kaburagi: J. Phys.: Condens. Matter {\bf 15} (2003) 5979.
\bibitem{sano} K. Sano and K. Takano: J. Phys. Soc. Jpn. {\bf 69} (2000) 2710.
\bibitem{kiku} H. Kikuchi, Y. Fujii, M. Chiba, S. Mitsudo, T. Idehara, T. Tonegawa, K. Okamoto, T. Sakai, T. Kuwai, and H. Ohta: Phys. Rev. Lett. {\bf 94} (2005) 227201.
\bibitem{ohta} H. Ohta, S. Okubo, T. Kamikawa, T. Kunimoto, Y. Inagaki, H. Kikuchi, T. Saito, M. Azuma, and M. Takano: J. Phys. Soc. Jpn. {\bf 72} (2003) 2464.
\bibitem{izuoka} 
A. Izuoka, M. Fukada, R. Kumai, M. Itakura, S. Hikami, 
and T. Sugawara: 
J. Am. Chem. Soc. {\bf 116} (1994) 2609. 
\bibitem{uedia} D. Uematsu and M. Sato: J. Phys. Soc. Jpn. {\bf 76} (2007) 084712.
\bibitem{dia4spin} N. B. Ivanov, J. Richter, and J. Schulenburg: Phys. Rev. B {\bf 79} (2009) 104412.
\bibitem{plaq} N.B. Ivanov and J. Richter: Phys. Lett. A {\bf 232} (1997) 308.
\bibitem{plaq2} J. Richter, N. B. Ivanov, and J. Schulenburg: J. Phys.: Condens. Matter {\bf 10} (1998) 3635.
\bibitem{koga1} A. Koga, K. Okunishi, and N. Kawakami: Phys. Rev. B {\bf  62} (2000) 5558.
\bibitem{koga2} A. Koga and N. Kawakami: Phys. Rev. B {\bf  65} (2002) 214415.
\bibitem{plaq3} J. Schulenburg and J. Richter: Phys. Rev. B {\bf 65} (2002) 054420.
\bibitem{plaq4} J. Schulenburg and J. Richter: Phys. Rev. B {\bf 66} (2002) 134419.
\bibitem{frulad1} T. Hakobyan, J. H. Hetherington, and M. Roger: Phys. Rev. B {\bf 63} (2001) 144433.
\bibitem{str1} L. \u{C}anov\`a, J. Stre\u{c}ka, and M. Jas\u{c}\u{u}r: J. Phys.: Condens. Matter {\bf 18} (2006) 4967.
\bibitem{str2} L. \u{C}anov\`a, J. Stre\u{c}ka, and T. Lu\u{c}ivjansk\'y: Condens. Matter Phys. {\bf 12} (2009) 353.
\bibitem{fuku} H. Kobayashi, Y. Fukumoto, and A. Oguchi: J. Phys. Soc. Jpn. {\bf 78} (2009) 074004.
\bibitem{m-d} C. Mathoni\`ere, J.-P. Sutter, and J. V. Yakhmi: in {\it Magnetism: Molecules to Materials IV,} ed. J. S. Miller and M. Drillon (Wiley, Weinheim, 2003) p. 1.
\bibitem{cb} Y. Hosokoshi and K. Inoue: in {\it Carbon Based Magnetism}, ed. T. L. Makarova and F. Palacio  (Elsevier B. V., Amsterdam, 2006) p. 107.
%\bibitem{aklt} I. Affleck, T. Kennedy, E. H. Lieb, and H. Tasaki: Commn. Math. Phys. {\bf 115} (1988) 477; Phys. Rev. Lett. {\bf 59} (1987) 799.
\bibitem{lm} %[11]
E. Lieb and D. Mattis: J. Math. Phys. {\bf 3}, (1962) 749. 
\bibitem{kene} T. Kennedy: J. Phys.: Condens. Matter {\bf 2} (1990) 5737.
\bibitem{ft} G. F\'ath and J. S\'olyom: Phys. Rev. B {\bf 44} (1991) 11836. 
\bibitem{wu} F. Y. Wu: Rev. Mod. Phys. {\bf 54} (1982) 235.
\bibitem{barbar} M. N. Barber: {\it Phase Transitions and Critical Phenomena 8}, ed. C. Domb and J. L. Lebowitz (Academic Press, London, 1983) p. 146.
\bibitem{takh} L.A. Takhtajan: Phys. Lett. {\bf 87A} (1982) 479.

\bibitem{babu} H. M. Babujian: Phys. Lett. {\bf 90A} (1982) 479.
\bibitem{ia}
I. Affleck and F. D. M. Haldane: Phys. Rev. {\bf B36} (1987) 5291.
\bibitem{yt}
Y. Kato and A. Tanaka: J. Phys. Soc. Jpn {\bf 66} (1997) 3944.
\bibitem{kn}
A. Kitazawa and K. Nomura: J. Phys. Soc. Jpn. {\bf 66} (1997) 3944.
\bibitem{ss} S. Sachdev and T. Senthil:   Ann. Phys. {\bf 251} (1996) 76.
\bibitem{bkk} L. Bartosch, M. Kollar, and P. Kopietz:  Phys. Rev. B {\bf 67} (2003) 092403.
\bibitem{ir} N. B. Ivanov and J. Richter:  Phys. Rev. B {\bf 69} (2004)  214420.
\bibitem{ym} S. Yoshikawa and S. Miyashita: J. Phys. Soc. Jpn. Suppl. {\bf 74} (2005) 71.
\bibitem{kh} K. Hida: J. Phys. Condens. Matter: {\bf 19}  (2007) 145225.
\bibitem{htsdec}
K. Hida and K. Takano: 
Phys. Rev. B {\bf 78} (2008) 064407.
\bibitem{filho} R. R. Montenegro-Filho and M. D. Coutinho-Filho: Phys. Rev. B {\bf 78} (2008) 014418. 
\bibitem{takahashi} M. Takahashi: Prog. Theor. Phys. Suppl. {\bf 87} (1986) 233.
\bibitem{yamamoto} S. Yamamoto: Phys. Rev. B {\bf 59} (1999) 1024.
\bibitem{yf} S. Yamamoto and T. Fukui: Phys. Rev. B {\bf 57} (1998) R14008.
\end{thebibliography}
\end{document}